# On the orientation of historic Christian churches of Fuerteventura: conciliating tradition, winds and topography

M.F. Muratore (1), A. Gangui (2), M. Urrutia-Aparicio (3), C. Cabrera (4), J.A. Belmonte (3)

(1) CONICET – Universidad Nacional de Luján and Instituto de Astronomía y Física del Espacio (IAFE), Argentina
(2) CONICET – Universidad de Buenos Aires and Instituto de Astronomía y Física del Espacio (IAFE), Argentina
(3) Instituto de Astrofísica de Canarias and Universidad de La Laguna, Spain
(4) Agrupación Astronómica de Fuerteventura, Fuerteventura, Spain

**Abstract**

We present the results of an analysis of the precise spatial orientation of colonial Christian churches located in the Canary Island of Fuerteventura (Spain). Our sample consists of 48 churches, most built during the period between the Castilian conquest led by the Norman Jean de Béthencourt in the 15th century and the end of the 19th century. We examine whether the standard tradition was followed regarding the orientation of the apses of historic churches eastwards. While most of the religious constructions in the sample have their main axes oriented within the solar range, the statistical analysis also reveals the presence of two different groups of churches with different possible interpretations. For the first group, mainly composed of churches located in the central part of the island, an anomalous tendency to orientate them towards a declination of c. –14º is detected. We provide some possible explanations for this, which include the date of a traditional Canarian celebration, an eventual imprint of topography, and the possibility of sunset orientations. Also, this particular value of declination is close to –16.3º, the declination of Sirius during the 17th century. Therefore, we provide ethnographic data that might support an eventually controversial 'bright star' orientation. For the second group, meanwhile, we find a pattern of orientation where the apse of the churches points slightly to the north of due east. We propose this might signal constructions that were oriented to the rising Sun on dates close to Easter, one of the most important festivities of Christianity.

## Introduction

Since its very beginnings, archaeoastronomy has concentrated its main efforts on the study of European megaliths, Egyptian pyramids, Mesoamerican temples, and other historical constructions, investigating the possible influences of celestial bodies on their designs (Ruggles 1999a; Krupp 1988; Aveni 2001). The study of the orientations of medieval churches has been another of the most popular objectives as well. On this last subject, recent work shows that the prescriptions for the orientation of churches' apses eastwards were very systematically followed throughout Europe during the Middle Ages (González-García 2015, and references therein).

From ancient texts, we know that the spatial orientation of historic Christian churches is one of the outstanding features of their architecture, with a notable tendency to orient the altars of the temples within the solar range (see, e.g., McCluskey 2015). Namely, the main axis of the church, from the narthex to the altar, should be aligned with the points on the horizon from where the Sun rises on different days of the year. Among these days, there is a marked preference for those corresponding to the equinoxes, even though this term may have different meanings (Ruggles 1999b; González-García and Belmonte 2006; Belmonte 2021).

Although researchers have mainly focused on analysing specific churches in the British Isles and continental Europe (see, e.g., Brady 2017), concentrating on their orientations and eventual illumination events, studies on the orientation of temples in periods after the Middle Ages and in regions far from the European centre have only gradually been developed. This late interest may be due to the widespread belief that the orientation of the churches lost its importance after the 16th century (Nissen 1906, 413; Arneitz *et al.* 2014). It is in this context that we place the present study. As we shall see, a large majority of the churches and chapels on the Canary Island of Fuerteventura began to be erected decades after the conquest and colonisation of the island by the Norman conquerors, who had the approval and support of the Crown of Castile at the very beginning of the 15th century (Cioranescu 1987).

Let us also note that one of the most direct influences for the construction of churches in colonial lands is the 16th century Council of Trent (1545–1563), after which Cardinal Charles Borromeo published his *Instructiones Fabricae et Supellectilis Ecclesiasticae* (1577), which was widely disseminated. In these instructions, in particular, Borromeo indicated the correct direction for the main altar:

> *Now, the placement of this altar must be elected at the head of the church, at the highest place where the main door is located; its back wall [of the church] points in a straight line towards the east, even though the people be placed at the back. And never orient it [the back wall] towards the solstitial east, but towards the equinoctial east* […]. (Borromeo 1985 [1577], 15, our translation)

These and other historical milestones, like for example the change from the Julian to the Gregorian calendar, could have influenced the traditions on churches' orientation during the long time elapsed since the construction of the earliest chapels on the island.

In a previous study, a preliminary discussion of our data has been offered. These had been collected in a fieldwork campaign in which orientation measurements, corresponding to the full sample of the historic (colonial and more recent eclectic styles) churches of Fuerteventura, were obtained (Muratore and Gangui 2020). The present paper further extends and completes the data analysis. This analysis complements previous works on the orientation of churches in the Canary Islands. Thus, it constitutes a follow up of our main project, which aims to investigate whether the texts of early Christian writers and apologists, regarding the orientation of religious architecture *ad orientem*, were – or were not – respected in this limited territory, located far from the European centres of power (Vogel 1962).

## The Castilian conquest and colonisation of Fuerteventura

The conquest of the easternmost Canary Islands began in 1402, was led by the Normans Jean de Béthencourt and Gadifer de La Salle and authorised by King Henry III of Castile. The chronicle of this process received the name of *Le Canarien* (see Aznar *et al.* 2003 for the most recent edition). They were accompanied by the secular priest Juan de Leverrier and the monk Pedro le Bontier, spiritual leaders of the expedition, who were the ones in charge of writing the chronicle as well as to start the process of Christianisation (Caballero Mújica and Riquelme Pérez 1999). After arriving and settling in Lanzarote, the expedition made continuous incursions into the neighbouring island of Fuerteventura. In 1404, Béthencourt and de La Salle founded Betancuria, which became Fuerteventura's first permanent settlement and was designated its capital. Béthencourt later returned to the continent to seek recognition and new support from the King of Castile. However, during his return, his ally Gadifer de La Salle decided to abandon the islands. Even so, within a few years, Fuerteventura was under control and the new society developed in a period of coexistence and mixing process between the European settlers and the natives, the *Maxos*.

From those early days, the main parish church on which the whole island depended was located in Betancuria. In the following decades, other urban centres were developed and populated and, slowly, the territory saw the emergence of estates and hamlets. In most of these villages, the growing population was accompanied by the construction of small Christian chapels and temples that reflected the new religious and social situation (Bethencourt Massieu 1973).

**Historic churches of Fuerteventura**

The first religious constructions on Fuerteventura were small, simple chapels with a single enclosure. One example is San Isidro Labrador, built in 1714 in the village of Triquivijate, in the municipality of Antigua. This church, like so many others, has a rectangular ground plan, a flat front façade and is now covered by a three-slope tile roof. Many of these chapels have more than one entrance door, such as San Isidro, whose side entrance is open towards the southern sector, and both the latter and the main doorway have openings made of light stonework, which are finished off with semi-circular arches (see Fig. 1).

Over the years, some of these constructions have had small chapels added to their chancel, sacristies on their sides, and other practical elements. San Isidro, for example, has a small bulrush (belfry) on the upper right side of its façade with a single opening, which serves as a modest bell tower, made of light-coloured stonework. On the left side, a small sacristy is attached to the exterior wall, where it adjoins the main nave.

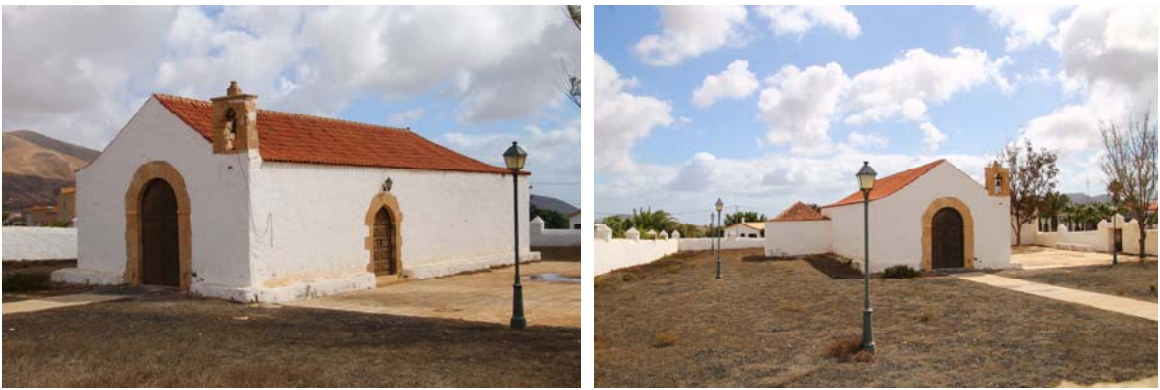

**FIGURE 1**. San Isidro Labrador in Triquivijate is a typical example of Christian chapel of Fuerteventura. A small sacristy connects to the main nave, whose façade has a modest bulrush that serves as a bell tower.

The small temple of San Isidro Labrador, like so many others built on Fuerteventura, was funded by the local people. Hence, many of them were not subjected to strict construction plans. For this reason, both their ground plan and their structure were built – and over the years they grew – according to the needs of the moment.

Over time, some churches were sufficiently enlarged to acquire a certain monumental character. They got portals with large semi-circular arches, bulrushes with one or more bays (as is the case of Santa Ana in Casillas del Ángel), and even bell towers (as at Nuestra Señora de Regla, in Pájara), with several bells and gabled or hipped roofs, in most cases tiled (Fig. 2).

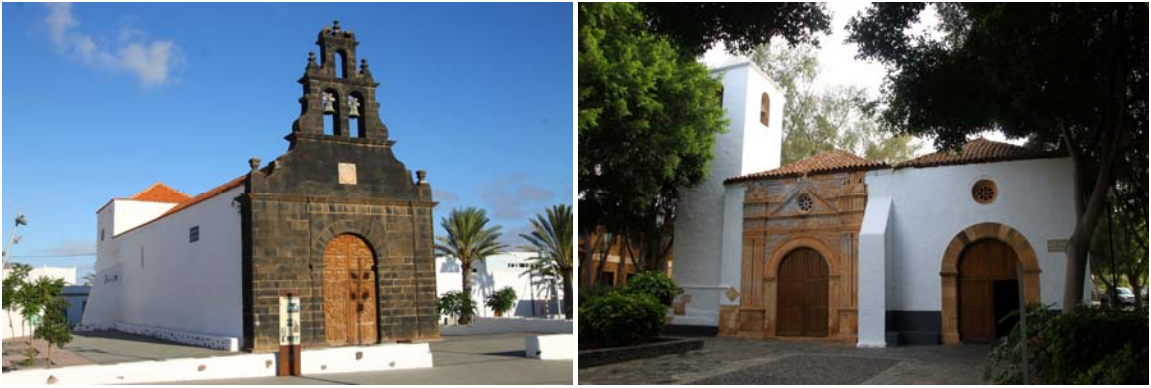

**FIGURE 2.** The churches of Santa Ana in Casillas del Ángel (left) and Nuestra Señora de Regla (Our Lady of Regla) in the village of Pájara (right) are amongst the largest and most elaborate religious constructions of the island.

On those occasions when the churches have not still been suffocated by the village growing around them, the enclosures remain surrounded by a wide outer crenellated wall (*muro almenado*) or barbican. There are nice examples of this at San Isidro (Fig. 3), at San Antonio de Padua in Lajares and at Nuestra Señora de Guadalupe in Agua de Bueyes, among others. In the case of the church of Agua de Bueyes, we know that the gates in its barbican were installed only in 1792 for the purpose of preventing the entry of livestock in the sacred precinct (Cerdeña Armas 1987).

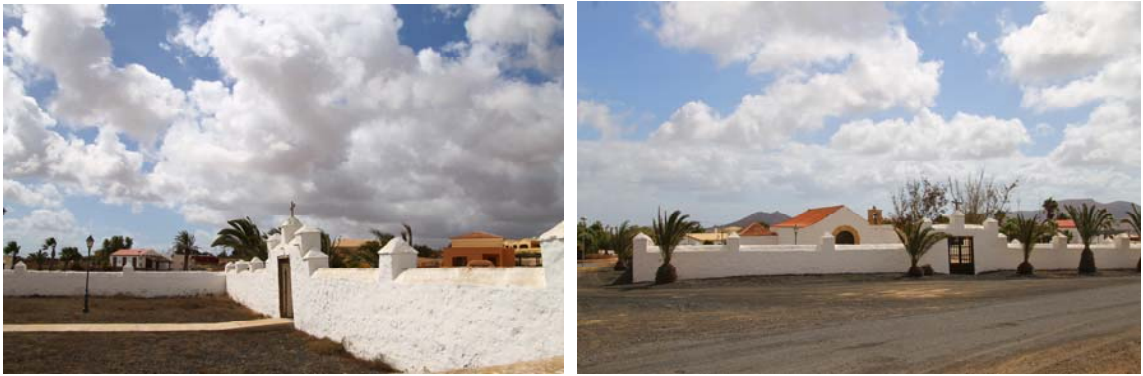

**FIGURE 3.** The ample barbican surrounding San Isidro Labrador has two entrances with wooden doors facing the entrances to the church building.

### Data sample and methods

Our present analysis is the continuation of a large-scale project carried out in the Iberian Peninsula and the Canary Islands. In the latter location, we have already focused on the precise orientation of the colonial churches on the islands of Lanzarote (Gangui *et al.* 2016a) and La Gomera (Di Paolo *et al.* 2020), as well as of those located in the city of San Cristobal de La Laguna, in the island of Tenerife (Gangui and Belmonte 2018).

In the present paper, we undertake the full systematic analysis of the orientation of the Christian religious constructions of Fuerteventura. Our main interest now is to conduct a statistical analysis of the sample, as this can provide us with unique archaeoastronomical data comprising a compact set of old churches whence we can search for Catholic religious traditions and even for pre-European ones (Belmonte 2015), including astronomical ones, or a mix of both. As with previous works, this may offer us a broader understanding of a key aspect of Canarian culture and traditions (Gangui *et al.* 2016b).

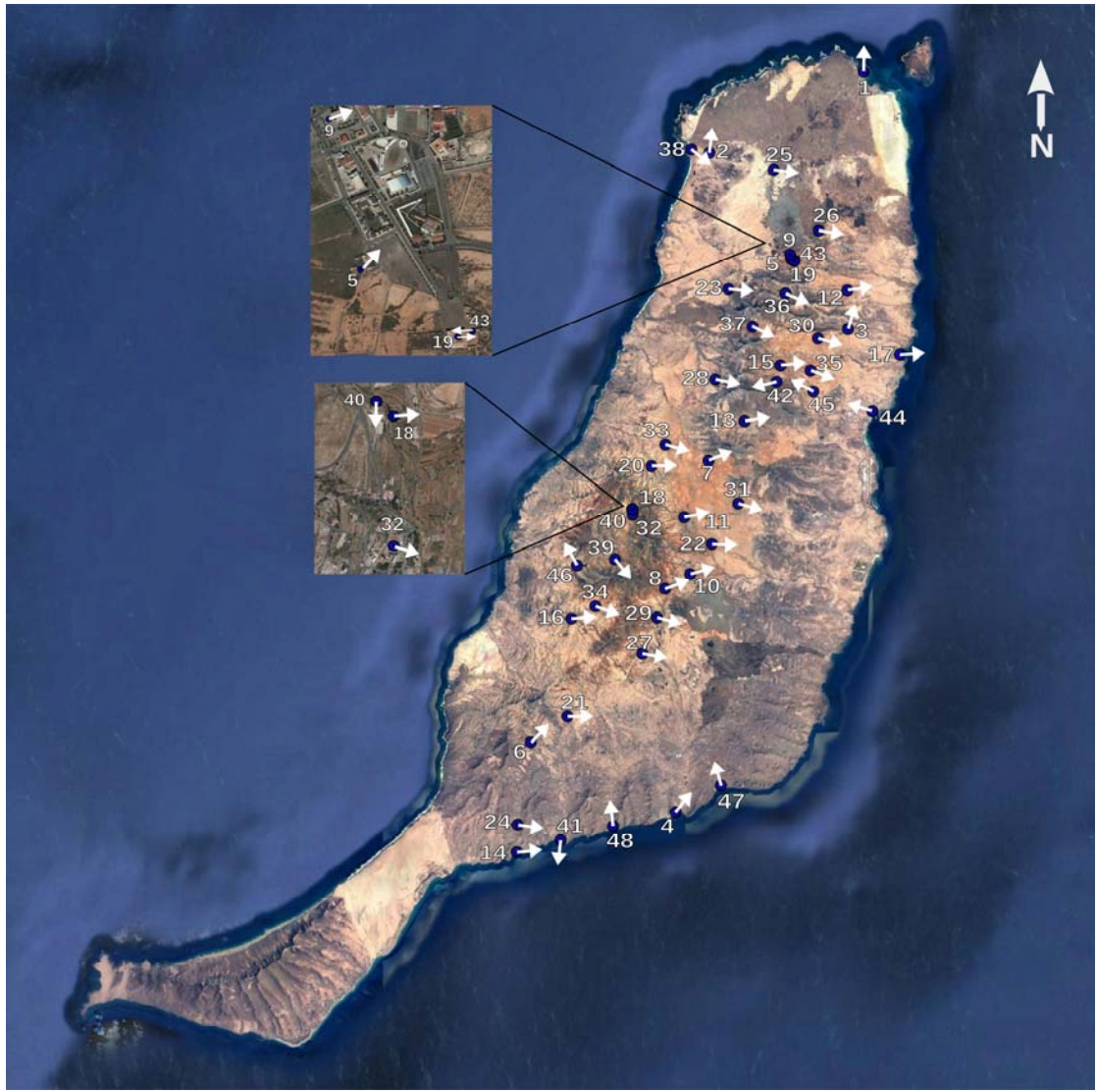

**FIGURE 4.** Map of the island of Fuerteventura with the geographical location of the full sample of churches measured (dark circles) and the orientation of the axis of the buildings in the direction of the altar (white arrows). Note the remarkable absence of churches in the Peninsula of Jandía to the south. This was an area of communal pasturage for centuries. Image on a map courtesy of Google Earth, adapted from Muratore and Gangui (2020).

Figure 4 shows the geographical location of all the churches and chapels measured (marked with dark circles, numbers according to Table 1), together with the orientation of the axis of the buildings in the direction of the altar (arrows, pointing according to the azimuths given in Table 1). Note that in the surroundings of the town of La Oliva, towards the north of the island, there are four constructions geographically very close to each other (the church of La Candelaria and three small chapels with dissimilar orientations), which explains the presence of overlapping circles there (we have enlarged the region in the inserted figure, including the four arrows). Something similar occurs in the Villa de Betancuria, in the central region, where the parish church, a convent and a chapel are all located a few hundred metres apart and have quite different orientations.

We obtained our measurements using a tandem instrument Suunto 360PC/360R, which incorporates a clinometer and a compass with a precision of 0.5°, and by analysing the surroundings (landscape) of each of the buildings. We then corrected the azimuth data according to the local magnetic declination (Natural Resources Canada 2015), getting values always within the range 4°14' to 4°21' W for different sites of the island. Our data is the result of several on-site measurements with a single instrument, taking the axes of the churches, from the back of the buildings towards the altars, as our main guide. In most cases, as only a tiny number of churches are today surrounded by modern buildings, we could also verify that the lateral walls were parallel to their axes.

Table 1 lists the identification of the churches, along with their geographical location and archaeoastronomical data: the measured azimuth and the angular height of the point of the horizon towards which the altar of the church is facing, as well as the derived computed declination corresponding to the central point of the solar disc (Ruggles 2015a; 2015b). The measured height of the horizon was appropriately corrected for atmospheric refraction (Schaefer 1993) and when the horizon was blocked, we employed the digital elevation model based on the Shuttle Radar Topographic Mission (SRTM) available at HeyWhatsThat (Kosowsky 2021), which gives angular heights within a 0.5° approximation. The uncertainty associated with each value of declination was obtained by error propagation (Ruggles 2015a), which gave a mean value of 0.7°. Orientation (azimuth) measurements are presented in Figure 5.

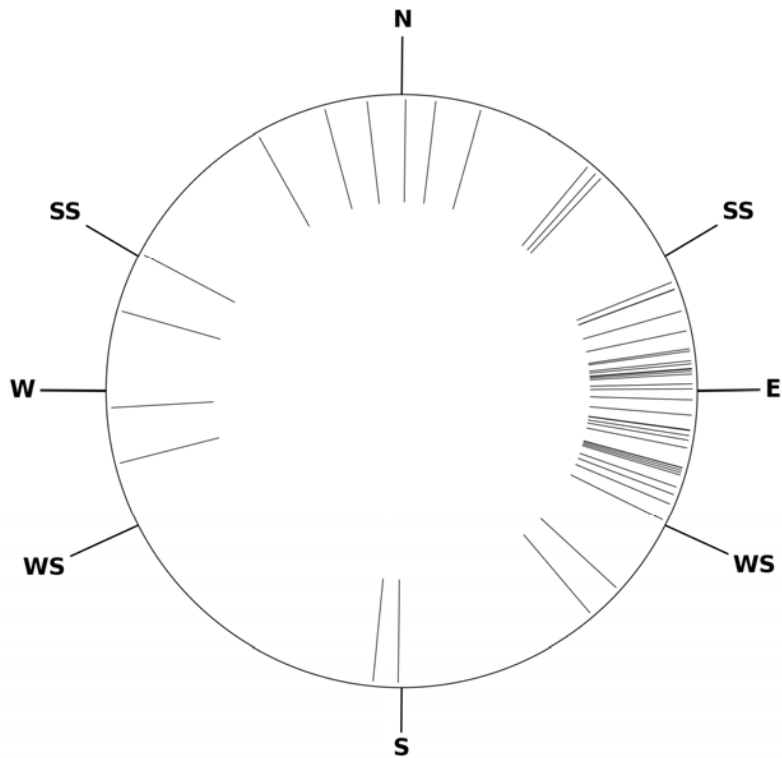

**FIGURE 5.** Orientation diagram for the churches and chapels of Fuerteventura, obtained from the azimuths of Table 1, considering the direction towards the altar. SS and WS mark the positions of the Sun when crossing the (ideal, flat) horizon during the northern-hemisphere summer and winter solstices. In the eastern quadrant, SS and WS correspond to azimuths 62.7° and 116.6°, respectively. This ideal horizon diagram is useful to illustrate general trends. Orientations within the solar range are predominant, but not exclusive. Interestingly, most of the anomalies to this rule are found with churches built in the 20th century. See text for further details.

Table 1: Orientations for the chapels and churches of Fuerteventura, ordered by increasing azimuth. For each building, we show the location, identification, the geographical latitude and longitude (L and l), the astronomical azimuth (a) taken along the axis of the building towards the apse, the horizon angular height (h) in that direction (including the correction due to atmospheric refraction; B means the horizon was blocked and we used HWT data), both rounded to $1/2^0$ approximation and expressed in decimal degrees, and the corresponding resultant declination (δ). The last column indicates the Patron saint date of each church and the Orientation (Gregorian date unless Julian is specified), which is computed by considering the documented year of the construction of each church and then estimating the dates when the declination of the Sun is the one indicated. The most likely date of construction is included together with the name in the second column, and we take it as the first mention of a church in that location or the date of its last major reform. The dates of those churches built over earlier structures have been marked with an asterisk. Finally, the numbers in the first column correspond to those signaling the geographical locations of the churches in the map of the island of Figure 4.

| Location | Name (date) | L (º, N) | l (º, W) | a (º) | h (º) | δ (º) | Patron saint date / Orientation |
|---|---|---|---|---|---|---|---|
| (1) Corralejo | Ntra. Sra. del Carmen (1992) | 28.7425 30 | 13.8672 66 | 1.0 | –0.6 | 60.7±0.5 | 16th Jul / —— |
| (2) El Roque | San Martín de Porres (c. 1985) | 28.6841 59 | 13.9942 20 | 6.5 | B 0.0 | 60.6±0.6 | 3rd Nov / —— |
| (3) Guisguey | San Pedro Apóstol (--) | 28.5578 82 | 13.8806 84 | 16.0 | +7.9 | 64.5±0.7 | 29th Jun / —— |
| (4) Gran Tarajal | Ntra. Sra. de la Candelaria (c. 1900) | 28.2120 27 | 14.0218 27 | 39.5 | B +2.2 | 44.1±0.7 | 2nd Feb / —— |
| (5) La Oliva | Ntra. Sra. De Puerto Rico (Capellanía) (c. 1500) | 28.6079 69 | 13.9274 10 | 42.0 | +2.2 | 42.3±0.7 | ?? / —— |
| (6) Cardón | Ermita de Cardón (c. 1980) | 28.2622 20 | 14.1405 38 | 43.0 | +7.4 | 44.3±0.7 | ?? / —— |
| (7) La Ampuyenta | San Pedro de Alcántara (c. 1681) | 28.4635 12 | 13.9952 26 | 68.0 | +6.4 | 22.2±0.7 | 19th Oct / 29th May - 15th Jul |
| (8) Agua de Bueyes | Ntra. Sra. de Guadalupe (1642) | 28.3721 39 | 14.0307 63 | 69.5 | +2.2 | 18.9±0.7 | 6th Sep / 13th May - 31st Jul |
| (9) La Oliva | Ntra. Sra. de la Candelaria (c. 1690) | 28.6111 31 | 13.9281 46 | 70.0 | +2.8 | 19.1±0.7 | 2nd Feb / 13th May - 30th Jul |
| (10) Valle de Ortega | San Roque (c. 1730) | 28.3825 75 | 14.0102 52 | 74.0 | +0.6 | 14.2±0.7 | 16th Aug / 27th Apr - 17th Aug |
| (11) Antigua | Ntra. Sra. de Antigua (c. 1550) | 28.4232 58 | 14.0151 06 | 78.0 | B +1.2 | 10.9±0.7 | 15th Aug / 7th Apr - 17th Aug (Julian date) |
| (12) La Caldereta | Ntra. Sra. de los Dolores (1796) | 28.5856 25 | 13.8812 28 | 82.0 | –0.6 | 6.9±0.7 | 15th Sep / 5th Apr - 6th Sep |
| (13) Casillas del Ángel | Santa Ana (1781) | 28.4918 43 | 13.9656 58 | 82.0 | +3.3 | 8.4±0.7 | 26th Jul / 10th April - 2nd Sep |
| (14) La Lajita | N. Sra. de la Inmaculada Concepción (c. 1995) | 28.1832 81 | 14.1518 36 | 84.0 | +4.8 | 7.4±0.7 | 8th Dec / 8th Apr - 6th Sep |
| (15) Tetir | Sto. Domingo de Guzmán (c. 1750) | 28.5318 75 | 13.9362 38 | 85.0 | –1.2 | 4.0±0.7 | 8th Aug / 29th Mar - 14th Sep |
| (16) Pájara | Ntra. Sra. de Regla (1653) | 28.3508 08 | 14.1074 16 | 85.5 | +3.8 | 5.6±0.7 | 8th Sep (?) / 2nd Apr - 10th Sep |
| (17) Puerto Lajas | Virgen del Pino (c. 1965) | 28.5397 75 | 13.8379 55 | 86.0 | –0.6 | 3.4±0.7 | 8th Sep (?) / 28th Mar - 16th Sep |
| (18) Betancuria | Iglesia Convento de San Buenaventura (1653*) | 28.4286 87 | 14.0572 93 | 86.0 | +12.9 | 9.4±0.7 | 15th Jul (?) / 12th Apr - 31st Aug |
| (19) La Oliva | Ermita interna, Casa de los coroneles (1742) | 28.6065 54 | 13.9250 92 | 87.0 | +0.6 | 3.1±0.7 | ?? / 27th Mar - 17th Sep |
| (20) Valle Santa Inés | Santa Inés (c. 1580) | 28.4596 03 | 14.0420 09 | 88.5 | +0.6 | 1.4±0.7 | 21st Jan / 13th Mar - 11th Sep (Julian date) |
| (21) Tesejerague | San José (c. 1725) | 28.2806 37 | 14.1104 90 | 89.5 | +10.4 | 5.2±0.7 | 19th Mar / 1st Apr - 11th Sep |
| (22) Las Pocetas | San Francisco Javier (1771) | 28.4039 53 | 13.9924 15 | 91.5 | +2.2 | –0.4±0.7 | 3rd Dec / 18th Mar - 26th Sep |
| (23) Tindaya | Ntra. Sra. de la Caridad (c. 1760*) | 28.5864 84 | 13.9788 57 | 94.5 | +6.9 | –0.8±0.7 | 15th Aug / 17th Mar - 27th Sep |
| (24) Tarajal de Sancho | Ntra. Sra. de Fátima (c. 1950) | 28.2030 85 | 14.1511 40 | 97.5 | +7.4 | –3.2±0.7 | 13th May / 12th Mar - 3rd Oct |
| (25) Lajares | San Antonio de Padua (c. 1750) | 28.6720 77 | 13.9414 69 | 98.0 | +0.6 | –6.5±0.7 | 13th Jun / 3rd Mar - 12th Oct |
| (26) Villaverde | San Vicente Ferrer (c. 1750) | 28.6281 85 | 13.9047 35 | 99.0 | +8.4 | –3.6±0.7 | 5th Apr / 10th Mar - 4th Oct |
| (27) Tuineje | San Miguel Arcángel (c. 1700) | 28.3255 41 | 14.0493 61 | 99.5 | B +1.2 | –8.0±0.7 | 29th Sep / 27th Feb - 15th Oct |
| (28) Tefía | San Agustín (1713) | 28.5220 16 | 13.9895 25 | 101.0 | +7.9 | –6.0±0.7 | 28th Aug / 4th Mar - 10th Oct |
| (29) Tiscamanita | San Marcos Evangelista (1699) | 28.3515 85 | 14.0372 00 | 105.0 | –0.6 | –13.6±0.7 | 25th Apr / 11th Feb - 31st Oct |
| (30) El Time | Ntra. Sra. de la Merced (1672) | 28.5509 14 | 13.9056 13 | 105.5 | –1.2 | –13.9±0.7 | 24th Sep / 10th Feb - 31 Oct |
| (31) Triquivijate | San Isidro Labrador (1713) | 28.4331 00 | 13.9708 22 | 105.5 | –0.6 | –14.1±0.7 | 15th May / 10th Feb - 2nd Nov |
| (32) Betancuria | Santa María (c. 1650*) | 28.4250 22 | 14.0572 85 | 106.0 | +7.9 | –10.2±0.7 | 8th Dec / 21st Feb - 21st Oct |
| (33) Llanos de la Concepción | Ntra. Sra. de la Inmaculada Concepción (1784) | 28.4752 94 | 14.0304 11 | 106.5 | +1.7 | –13.8±0.7 | 8th Dec / 11th Feb - 31st Oct |
| (34) Toto | San Antonio de Padua (c. 1775) | 28.3595 78 | 14.0877 36 | 109.0 | +2.2 | –15.7±0.7 | 13th Jun / 5th Feb - 7th Nov |
| (35) Los Estancos | Santa Rita (c. 1980) | 28.5282 29 | 13.9116 51 | 111.0 | –0.6 | –18.4±0.7 | 22nd May / 27th Jan - 17th Nov |
| (36) Vallebrón | San Juan Bautista (1704) | 28.5831 80 | 13.9320 58 | 113.0 | +3.8 | –17.9±0.7 | 24th Jun / 29th Jan - 15th Nov |
| (37) La Matilla | Ntra. Sra. del Socorro (c. 1716) | 28.5593 90 | 13.9592 46 | 116.0 | +1.2 | –22.2±0.7 | 8th Sep / 7th Jan - 6th Dec |
| (38) El Cotillo | Ntra. Sra. del Buen Viaje (c. 1690) | 28.6866 80 | 14.0090 98 | 132.5 | +1.7 | –35.3±0.7 | End of Aug / —— |
| (39) Vega de Río Palmas | Ntra. Sra. de la Peña (c. 18th cent.*) | 28.3932 20 | 14.0716 90 | 139.5 | +10.4 | –35.0±0.7 | 3rd Saturday Sep (?) / —— |
| (40) Betancuria | San Diego de Alcalá (1653) | 28.4290 98 | 14.0577 83 | 180.5 | +2.2 | –59.3±0.5 | 13th Nov / —— |
| (41) Tarajalejo | Ermita de Tarajalejo (c. 1995) | 28.1919 68 | 14.1159 93 | 185.5 | –0.6 | –61.9±0.6 | 2nd Sunday May / —— |
| (42) Tetir | San Andrés (1989) | 28.5201 50 | 13.9395 39 | 255.5 | +5.9 | –9.6±0.7 | 30th Nov / 22nd Feb - 20th Oct |
| (43) La Oliva | Ermita exterior, Casa de los coroneles (c. 1550 ?) | 28.6066 62 | 13.9247 26 | 267.0 | –0.6 | –3.1±0.7 | ?? / 2nd Mar - 23rd Sep (Julian date) |
| (44) Pto. del Rosario | Ntra. Sra. del Rosario (1824) | 28.4989 75 | 13.8605 17 | 286.0 | +2.8 | 15.2±0.7 | 3rd Oct / 30th Apr - 14 Aug |
| (45) La Asomada | Ntra. Sra. de Fátima (c. 1995) | 28.5130 17 | 13.9092 87 | 298.0 | +1.2 | 24.7±0.7 | 13th May / 'solstitial' |
| (46) Bco. de las Peñitas | Ermita del Malpaso (1748*) | 28.3888 33 | 14.1028 47 | 330.5 | +38.5 | 63.7±0.5 | 3rd Saturday Sep (?) / —— |
| (47) Las Playitas | San Pedro Apóstol (c. 1995) | 28.2309 54 | 13.9848 37 | 344.5 | +1.7 | 59.7±0.7 | 29th Jun / —— |
| (48) Giniginamar | Ntra. Sra. del Carmen (1992) | 28.2019 98 | 14.0731 44 | 353.0 | +3.3 | 64.2±0.6 | 16th Jul / —— |

For the analysis, we computed declination curvigrams that provide information on the probability of finding a particular value of the variable within our sample, following the methodology of several preceding works (González-García and Sprajc 2016; Urrutia-Aparicio *et al.* 2021a). For this, we used an appropriate smoothing of the histogram by a function called "kernel" to generate the kernel density estimation (KDE). For each value of the declination in Table 1, we multiplied the value of the number of occurrences by an Epanechnikov kernel function with a given passband or bandwidth, fixed to twice our mean error.

In order to know if a concentration of declination values is significant, we compared the distribution of our measurements versus the result expected from two possible distributions (see Fig. 6). The first one is a uniform distribution, in which the orientations are homogeneously shared among all possible azimuth values, with a flat horizon and a mean latitude of 28.5°, corresponding to Puerto del Rosario, the modern capital of Fuerteventura. The second one is a "solar" distribution, which represents the changing declination of the Sun over a flat horizon over one solar year (Waugh 1975, 9-12, 206).

We then scaled our curvigram with respect to one of those distributions to check whether our actual data departs significantly from them. We extracted a hundred random subsamples of a size equal to our measured data, from the uniform or solar distributions, and obtained the deviation of each of them to calculate the mean value. Afterwards, the uniform or solar distribution was subtracted from our measured one, and the resulting distribution was divided by the mean of the deviations.

We used the uniform distribution for the normalisation, since those peaks outside the solar range could be explained by a random pattern. Figure 6 shows the 100 random distributions obtained from the uniform distribution and with the same number of elements as our empirical data. This same methodology can be studied in detail in a very recent paper by González-García *et al.* (2021) on the archaeological site of Caral.

Interestingly, in the last two years, Silva (2020) and Abril (2021) have proposed some updates to the methodology that had been used in the last decade, including the one used in this work. The innovative approach of the former features a coordinate transformation that includes detailed horizon profile information. This is reflected in a more accurate sample declination distribution, provided that these horizon data were available for every place where measurements had been obtained. In our case, this data was not collected because our experience has shown that we do not expect to find significant changes in the height of the horizon for the azimuth and angular height instrumental accuracy under consideration. This is specially the case of Fuerteventura, where landscape presents very few abrupt features. The uncertainties, and therefore the bandwidth of the kernel, appear to be sufficient for the level of precision we are looking for when interpreting the patterns. Certainly, Silva's model can be really useful to carefully analyse single sites within a variegated landscape.

Returning to the normalisation process, it is worth mentioning that using this "normalised relative frequency" is equivalent to comparing our data with the results of a uniform or solar distribution of the same size as our data sample, and with a mean value equal to the mean value of our data. The scale is given by the standard deviation ($\sigma$), so we will consider that any peak rising above $3\sigma$ is statistically significant according to our analytical approach, as will be shown in Fig. 8 (below), where the data were normalised by the homogeneous distribution, considered more appropriate for our case.

Additionally, we conducted several Kolmogorov-Smirnoff tests that provide the level of confidence with which we can discard our null hypothesis, that is, that our orientation sample derives from the same distribution as a random one. We repeated this test for a hundred random samples extracted from the two distributions and, in both cases, the tests indicated that we cannot discard the null hypothesis with confidence. This seems to agree with Figure 6: orientations that fall outside the solar range can match a random sample obtained from a uniform distribution, and a few orientations within the solar range could be following the solar distribution.

It should be noted that some maxima that we have considered significant (and therefore subject of study) might have the same relative frequency when compared to an individual random distribution, such as any of those in Figure 6. This could lead to doubts about their significance. However, these maxima come from real, empirical, data without instrumental or any other kind of noise, with only the intrinsic error of the instruments and human precision. This implies that, although the accumulation peaks within the solar range could be seen as random and non-significant, this does not mean that they should lack a reasonable cultural explanation. Hence, a discussion on those maxima above the $3\sigma$ level in Figure 8 would indeed be worthy.

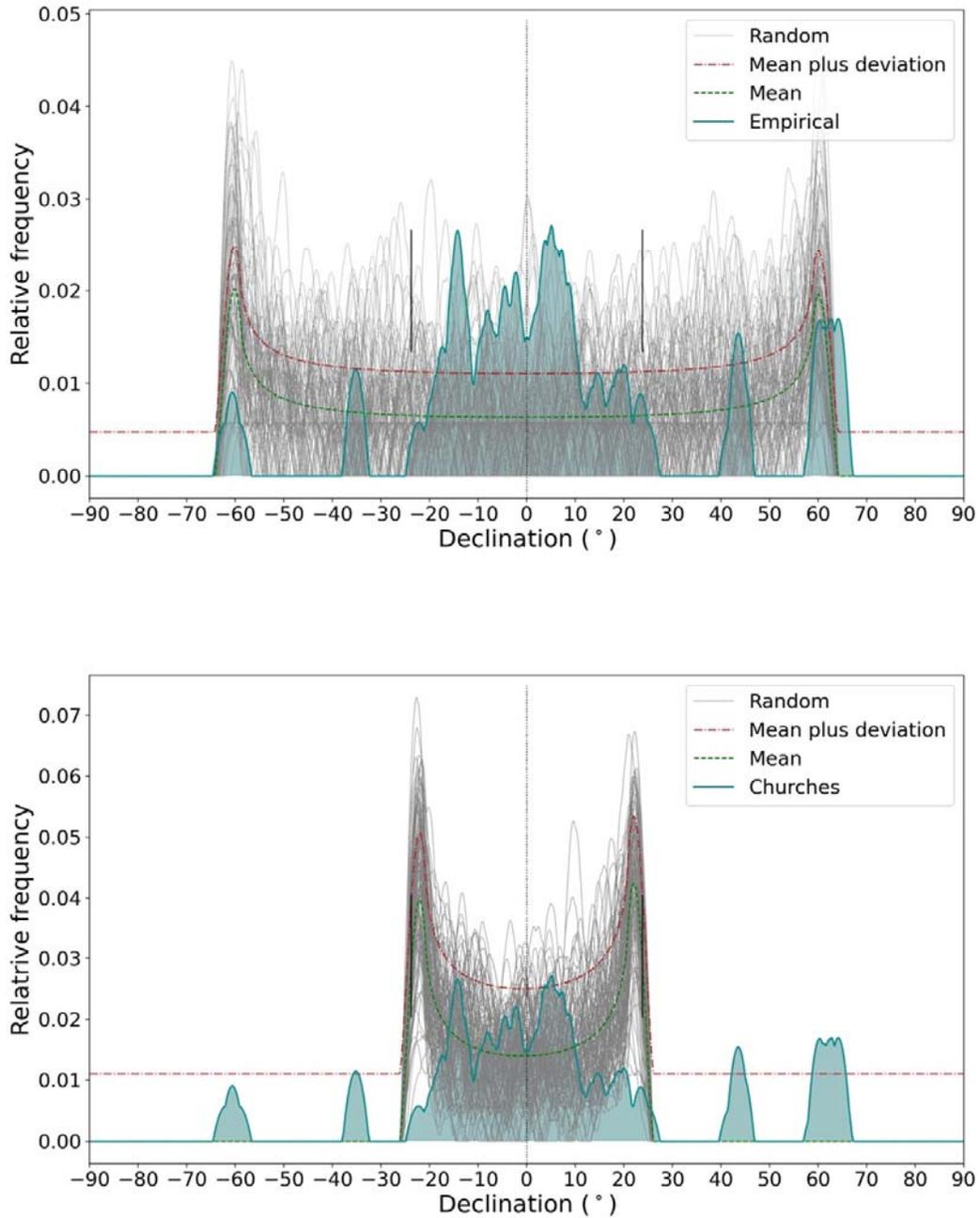

**FIGURE 6.** Declination histograms of our sample (cerulean colour, solid line) compared to: (up) a series of random distributions from a uniform interval (light grey, solid line), obtained by equally distributed azimuths from 0° to 359°, a flat horizon and a latitude of 28.5°, the mean and the mean plus the standard deviation are represented in green and red dashed lines; (bottom) a series of random distributions within the solar range (light grey, solid line), the mean of which represents the declination values of the Sun over a year (green, dashed line), and finally, the mean plus the standard deviation for this case (red, dashed line). The astronomical equinox is indicated by a vertical dotted line at a declination of 0º, while short solid vertical lines represent the winter and summer solstices. Note that in the top curvigram the churches oriented out of the solar range fall within the uniform distribution, whilst several peaks around the equinox might be significant. However, when the analysis is restricted to the solar distribution, this significance is nuanced. See the text for further discussion.

**Results**

Figure 5 shows the orientation diagram for the churches and chapels measured in Fuerteventura. Our sample is representative of the whole island, so it should allow us to infer general orientation patterns.

The dominant orientation pattern is an expected one within the solar range, according to the texts of early Christian writers and apologists (McCluskey 2015; Gangui *et al.* 2016a). However, a few small and several modern chapels are oriented within the northern and southern sectors, and therefore fall outside of the solar arc. Indeed, the great majority of the religious constructions of the island follow the canonical pattern, orienting their axes between the extreme azimuths (i.e., SS–WS) of the annual movement of the Sun as it crosses the local horizon. There are 34 churches, which is around 70% of the sample (91% if buildings prior to the 20th century are considered), whose orientations fall within the solar range, either to the east or to the west.

As shown in Figure 5, of the 48 measured orientations, 9 are directed towards the northern quadrant (azimuths between 315° and 45°), 3 towards the southern, 4 towards the western and the great majority (32) point towards the eastern quadrant. This result, although consistent with the pattern of orientations of groups of churches from earlier periods typical of the places of origin of the colonisers (Vogel 1962; González-García and Belmonte 2015), differs notably from the outcomes obtained in other islands of the archipelago. Interestingly, it substantially differs from the results from nearby Lanzarote (Gangui *et al.* 2016a), where a notable proportion of the churches were oriented approximately north-northeast (with a leeward entrance) to avoid the prevailing trade winds of the place.

It also differs from what was found in La Gomera (Di Paolo *et al.* 2020), where a roughly similar orientation pattern was found as in Lanzarote, but for a different reason. It was shown that in La Gomera the accumulation of orientations in the northeast sector is due to the orographic characteristics of the island, where several groups of temples are oriented following the direction of a couple of deep valleys among those that mark the island's "abrupt geography" (Díaz Padilla 2005) and that run approximately from southwest to northeast.

Now let us consider different groups of churches mainly according to their common broad azimuthal orientation (we will complete this discussion below, when we discuss the precise declination values of the different constructions). Two churches present orientations close to due east: the chapels of San José in Tesejerague (number 21 in Table 1; first half of the 18th century) and of San Francisco Javier in Las Pocetas (no. 22; from 1771; cf. Cerdeña Armas 1987). Other constructions with eastward orientations are, for example, Santa María de Betancuria (no. 32), the parish church of the Villa of the same name and, for decades after the conquest, of the whole of Fuerteventura, whose construction was commissioned by Jean de Béthencourt himself around 1410. Its main axis is oriented with an azimuth of 106º, which places it a little more than ten degrees to the south of due east. In the same capital city, although of a slightly later date, we find the church of the Franciscan Convent of San Buenaventura (no. 18), of which only the solid structure and walls remain from a 17th century reconstruction after the Berber attacks, as its roof did not survive a later fire. This convent church structure also faces east with an azimuth of 86º and is therefore just a few degrees north of due east. Finally, we find the church of Santa Ana, in Casillas del Ángel (no. 13, Fig. 2), whose axis is oriented to an azimuth of 82º and, in the town of Lajares, the chapel of San Antonio de Padua (no. 25), with an azimuth of 98º. Both buildings date from the latter part of the 18th century (Concepción Rodríguez 1989) and are less than 10° away from the geographical east.

There are also two churches that could be called approximately "solstitial", within our uncertainties, one historical and the other modern. Nuestra Señora del Socorro, in La Matilla (no. 37), which dates from 1716 (Cerdeña Armas 1987), has an axis that could be aligned with the direction of sunrise during the winter solstice (with an approximate azimuth of 116º and declination of −22.2º). In contrast, the more recent Nuestra Señora de Fátima at La Asomada (no. 45) follows the direction of sunset during the June solstice (with an azimuth of 298º and declination of 24.7º). However, it could also be oriented to sunrise at the December solstice if, instead of considering the direction towards the altar, we take the opposite, as has been seen on nearby islands (Gangui and Belmonte 2018).

Regarding the dozen churches in Table 1 that are oriented in the northern and southern sectors, and therefore far from the solar range, we believe that, for many of them, there are practical reasons that could explain their measured orientations. It could be due to their origins as private chapels belonging to high-status families in Fuerteventura as recorded in original documents (such as "Libros de Fábrica" or "Libros de Cuentas" -- Bonnet Reverón 1942; Bethencourt Massieu 1973; Cerdeña Armas 1987), or because their locations faced the seashore. Another practical reason could respond to the fact that some churches were oriented in a direction opposite to the dominant trade winds, as in Lanzarote (just think of the name of the island, whose Castilian meaning could be "Strong Winds").

Among the chapels whose orientations are distributed within the northern sector, there are several located in the southern and coastal region of the island, all of them built throughout the 20th century. Such is the case of the chapels of Virgen del Carmen, in Giniginamar (no. 48), and San Pedro Apóstol, in Las Playitas (no. 47). Both are of modern construction and open their doors leeward towards the sea, and it is therefore natural that their altars face north. The churches of San Martín de Porres, in El Roque (no. 2), or the chapel of Cardón (no. 6), both oriented in a northerly direction, are also of recent construction, and therefore fall into this category (also avoiding north winds) without much doubt.

The chapel at Tarajalejo (no. 41), however, although also located on the south coast and not far from Las Playitas, surprisingly has its chancel (and not its door) facing the coast, which is why its measured azimuth is a few degrees from due south. Another modern church in the south of the island is Nuestra Señora de la Candelaria in Gran Tarajal (no. 4). Its axis is oriented to the northeast and outside of the solar range; almost certainly, its builders did not consider the religious architectural tradition when building this large church in a fast-growing town.

In our sample, there are also a couple of chapels, built by private individuals, which deviate from the canonical orientation. This is the case for the chapel of La Capellanía in La Oliva (no. 5) which dates from around 1500 and is thought to have been the former home of a local clergyman. The chapel of San Diego de Alcalá in the village of Betancuria (no. 40), also falls into the category. Tradition has it that it was erected on the site of a small cave, a frequent place of prayer for the saint in the mid-15th century. Its orientation is almost along the meridian and very distant from the solar range.

In Vega de Río Palmas and its continuation in Barranco de las Peñitas there are two churches that depart from the canonical orientation tradition and that have a shared history (Hernández González 1990). The tiny chapel called Ermita del Malpaso (no. 46) dates from the 15th century and, as its name indicates, is very difficult to access. There remained for some time the sculpture of the Virgen de la Peña, the Patroness of Fuerteventura. Tradition says that it had been brought to Betancuria from France by Béthencourt but it was then hidden in Las Peñitas to save it from the Berber pirate attacks of the late 16th century. Due to the difficulty of accessing this chapel, the church of Nuestra Señora de la Peña (no. 39) was built in the 18th century in Vega de Río Palmas, for the convenience of the worshippers (Fig. 7).

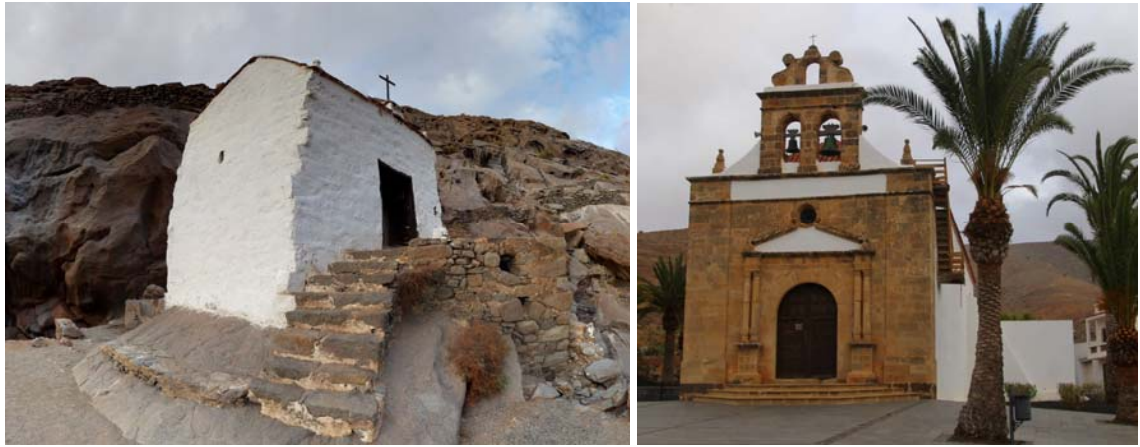

**FIGURE 7.** Ermita del Malpaso, located in the ravine of Las Peñitas, a tiny chapel of very difficult access (left image), and the church of Nuestra Señora de la Peña (right), built sometime later in the nearby village Vega de Río Palmas, a few kilometres from the former. In both cases their main axis is oriented outside the solar range, but the gate of the hermitage is open eastwards.

The axes of these two constructions deviate greatly from the canonical orientations. In the case of the small chapel, a close inspection of the surrounding landscape suggests that the movement of the Sun and the arc of the horizon where it rises and sets during the year were irrelevant to its builders (as the high value of "h" in Table 1 suggests). It was built on the slope of a deep secluded ravine, presumably for protection. Thus, while it is simple to understand that there was not much choice in the orientation of the chapel of Las Peñitas, the southeastern and therefore out of the solar range orientation of the later church of Nuestra Señora de la Peña is not adequately documented and is a subject that requires further exploration.

To better understand what we have discussed so far, Figure 8 presents a declination histogram (or curvigram) for the sample of churches, but now with the normalised relative frequency, which includes geographical location and local topography, and which enables another view of the concentration or the probability density of certain orientation patterns that might be a matter of interest within our sample.

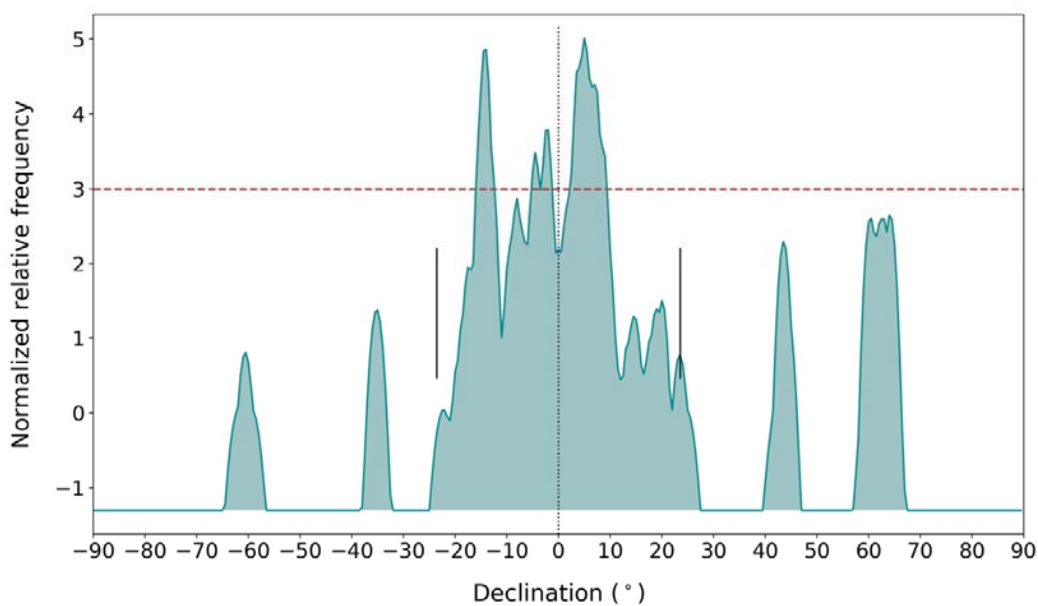

**FIGURE 8.** Normalised declination curvigram of the churches of Fuerteventura. The astronomical equinox is marked with a vertical dotted line, whereas the winter and summer solstices have shorter vertical solid lines. The horizontal red dashed line representing the 3σ level is a direct outcome of our normalisation process. Peaks rising above this line can be considered potentially significant.

The normalised curvigram shows a preference for orientations within the solar range that agrees with the azimuth values in Figure 5. These declination values can be associated with certain dates in the calendar that have been obtained from the JPL Ephemeris software available at the corresponding website (Horizons 2021). As most of the measured churches were built after the calendar reform in 1582, dates in Table 1 are given in the Gregorian calendar. The expected uncertainties depend on the time of the year and can be of about a day near the equinoxes but might increase considerably close to the solstices. Those orientations out of the solar range, of course, have not been translated into dates. The principal maximum in Figure 8 appears at a declination of c. 5º, which is around 1 April or 12 September. The second maximum, almost as significant as the former, is located at c. –14º, corresponding to around 10 February or 2 November. Finally, there are two consecutive peaks at c. –5º (around 8 March or 7 October) and c. –2º (around 14 March or 1 October).

### Discussion

Both the orientation diagram and the declination curvigram discussed in the previous sections imply that most of the orientations of Fuerteventura churches fit within the solar arc. Conspicuously, as we also pointed out, the great majority of the constructions' altars face the direction of sunrise at some time in the year. This orientation conforms with Christian tradition (McCluskey 2015) and, therefore, it could suggest that other more prosaic (say, topographic, climatic, etc.) explanations for variations are not at work here, as they were in Lanzarote and La Gomera.

However, it is important to note that out of the 34 churches whose orientations fall in the solar range, 26 (76%) do have a lateral door opening towards the south. The use of these doors to access the interior of the buildings would prevent the annoying prevailing northern trade winds of the region from disturbing the smooth running of the ecclesiastical activities. This is in contrast to the churches of Lanzarote (Gangui *et al.* 2016a), where the same strong winds are dominant and churches were often oriented towards north-northeast or were constructed with large barbicans, accordingly. We know that nowadays the local people of Fuerteventura frequently enter and exit through the side doors of the churches. So, this combination of canonical orientation plus a gate open southwards would permit both to save the religious tradition and to avoid the deterrent northern winds, a nice combination of orthodox and practical behaviours.

In Figure 8, there is a prominent maximum at a declination c. –14±2º which, a priori, has no obvious reason to stand out from the rest of the declination measurements that fit within the solar arc. The usual theory of orientations toward the sunrise on the church patron's feast day was discarded by individual evaluation, as contrasted and demonstrated in Table 1, in which no patron saint's day fits with the corresponding astronomical orientation. Other possible explanations, such as use of a magnetic compass to determine the axis of the church, or the avoidance of annoying winds, have also been tested and were rejected.

This is indeed an atypical orientation that differs greatly from what would be expected in the Iberian Peninsula (e.g., González-García and Belmonte 2015; Urrutia-Aparicio *et al.* 2021b). However, in most recent analysis of contrast data in the area of the Way of Saint James, such a peculiar southern behavior has also been identified (Urrutia-Aparicio *et al.* 2022). These authors have suggested a possible orientation towards Easter sunset that might be applied to this case.

At present we can think of another three possible explanations to try and justify this peak of approximately –14º in declination, but unfortunately none of them is conclusive: the traditional Canarian celebration of "Los Finaos"[1] (the local version of All Souls' Day or the Day of the Dead), a topographic orientation for churches located in a few valleys pointing towards a direction close to the 105º azimuth and, lastly, an unusual but appealing orientation to a bright star which might be supported by some relevant ethnographic testimonies. Below, we will briefly mention the first two examples, as this is not the first time that these possibilities appear in the literature on the Canary Island's church orientation. This is the case of La Laguna and the feast of San Cristóbal (Gangui and Belmonte 2018) and the topographic orientation of historical churches in La Gomera (Di Paolo *et al.* 2020). Then, we will delve with full more detail into the last, stellar, orientation pattern, which we consider more speculative than the previous two, but eventually more interesting due to its ethnoastronomical connections.

The celebration of "Los Finaos" (*los finados,* the dead) in Fuerteventura, and in general in all the Canary Islands, has always been a well-known occasion for the families to get together and remember their dead loved ones. The reunion takes place on the night of 1 to 2 November each year, during which friends and family gather for a vigil. If we consider churches number 29, 30, 31 and 33 in the last column of Table 1, we see that all of them have orientations compatible with this celebration date, matching approximately the –14º peak in declination. So, one cannot a priori exclude this circumstantial influence in our declination plot.

As regards the particular topography of some regions in Fuerteventura, one can check that, out of the same four above-mentioned churches in Table 1, the first three of them are located in valleys that descend to the seashore in a direction close to 105º in azimuth and with very low, actually negative, horizon elevation. As we can see in the penultimate column of Table 1, this also matches the –14º value in declination, so this "topographic forcing" could very well influence our declination plot and should not be excluded from the analysis.

Finally, let us consider our last possible explanation. As we mentioned, one unusual hypothesis would be to assign a non-solar origin, perhaps prior to the Christianisation of the region, say, to a brilliant planet or a bright star. In this sense, the *Gañanera* (the local name of Sirius), which in the 17th century had a declination of c. –16.3º, is perhaps an interesting possible target as we shall discuss. However, it is worth noting that there are no precedents in the literature which could relate the orientation of churches with this or any other star.

---

[1] Also called *Ranchos de Ánimas* in Fuerteventura.

The *Gañanera*[2] is well known in Fuerteventura as the best guide for agricultural activities, specifically, in the central area of the island where several of the churches of this group of orientations are located (see the data of Table 1 and Figure 4). If, for example, we consider ethnographic data, there are particular statements of old farmers and goatherds mentioning a bright and striking "star" which stands out among all the others:

> *Aquí por el sur sale una que nosotros le decimos la Gañanera. Sale de madrugada, de medianoche para el día e incluso antes [...] estos de la labranza se levantaban temprano, de Aguabueyes, y decían que cuando va bajita, dicen que es tiempo bueno para cuestión como este año a proponer, este año va baja, se levanta por allí (hacia el mar) y va caminando y se mete por allí, por los corrales aquellos (en la montaña hacia el sur) [...]. Ellos de allí no lo ven y entonces se levantaban de madrugá a venir a la montaña esta que le dicen Pedriales e iban a observarla allí.[3] Sale de medianoche en adelante. Incluso desde la cama la veo yo. Me despierto y miro para allí (hacia el SE) y la veo.[4]* (Belmonte and Sanz de Lara 2020, 83).

This statement by Victoriano Pérez (78 years old in 1996), from the coast village of Pozo Negro, may give us a hint that exploring the archaeoastronomical data of the church Nuestra Señora de Guadalupe (no. 8 in Table 1), located in Agua de Bueyes (*Aguabueyes* in Pérez's speech), might be a good idea. The church's declination in the direction of the altar is approximately +19°. However, if we take the opposite direction, towards the narthex (the main door), and estimate the height of the horizon with HeyWhatsThat (this datum was not taken on site), we readily obtain a value of $\delta = -14.5 \pm 0.7°$, close to the declination of Sirius during the 17th century (recall that the date of construction of the church is around 1642).

Additionally, this church has two clear "astral" symbols on the façade of the barbican main entrance (Fig. 9): an eight-pointed star, which is sometimes associated with Venus, and a six-pointed rosette, which is usually associated with the Sun. On many occasions, in the case of the central area of Fuerteventura, where many of these churches were built, the *Gañanera* played and still plays a role similar to Venus as a guidance for agricultural practices and could hence be confused or intermingled with this planet in the peasants' mind. However, we must admit that this is circumstantial evidence.

---

[2] *Gañán* is a Castilian Word for the one who drives the plough. In Fuerteventura, it received the typical suffix –*era*. In the sky, it followed the Plough as represented in Orion's Belt (Belmonte and Sanz de Lara 2020, figure 16).

[3] The Gañanera was so important that the farmers of Agua de Bueyes woke up early in the 'madrugá' (i.e., before dawn) to go up to a mountain (Pedriales) to observe it over the sea when it rises, before going to plough. The equivalence with Sirius was precisely confirmed because this is the only bright star which is seen in the adequate direction towards Pozo Negro.

[4] A free translation of a difficult peasant's language could be: *Here in the south comes one that we call the Gañanera. It leaves at dawn, at midnight towards the day and even before [...] those who did the ploughing got up early, at Aguabueyes, and they said that when it goes at a low level, they say that it is a good time for the question [of ploughing]. Like this year, they propose, ... this year it goes low, it gets up over there (towards the sea) and walks and hides through there, through those corrals (in the mountains to the south) [...]. They do not see it from there and then they would get up early to come to the mountain that they call Pedriales and they would go there to observe it. It rises from midnight onwards. I can even see her from bed. I wake up and look over there (to the SE) and I see it.*

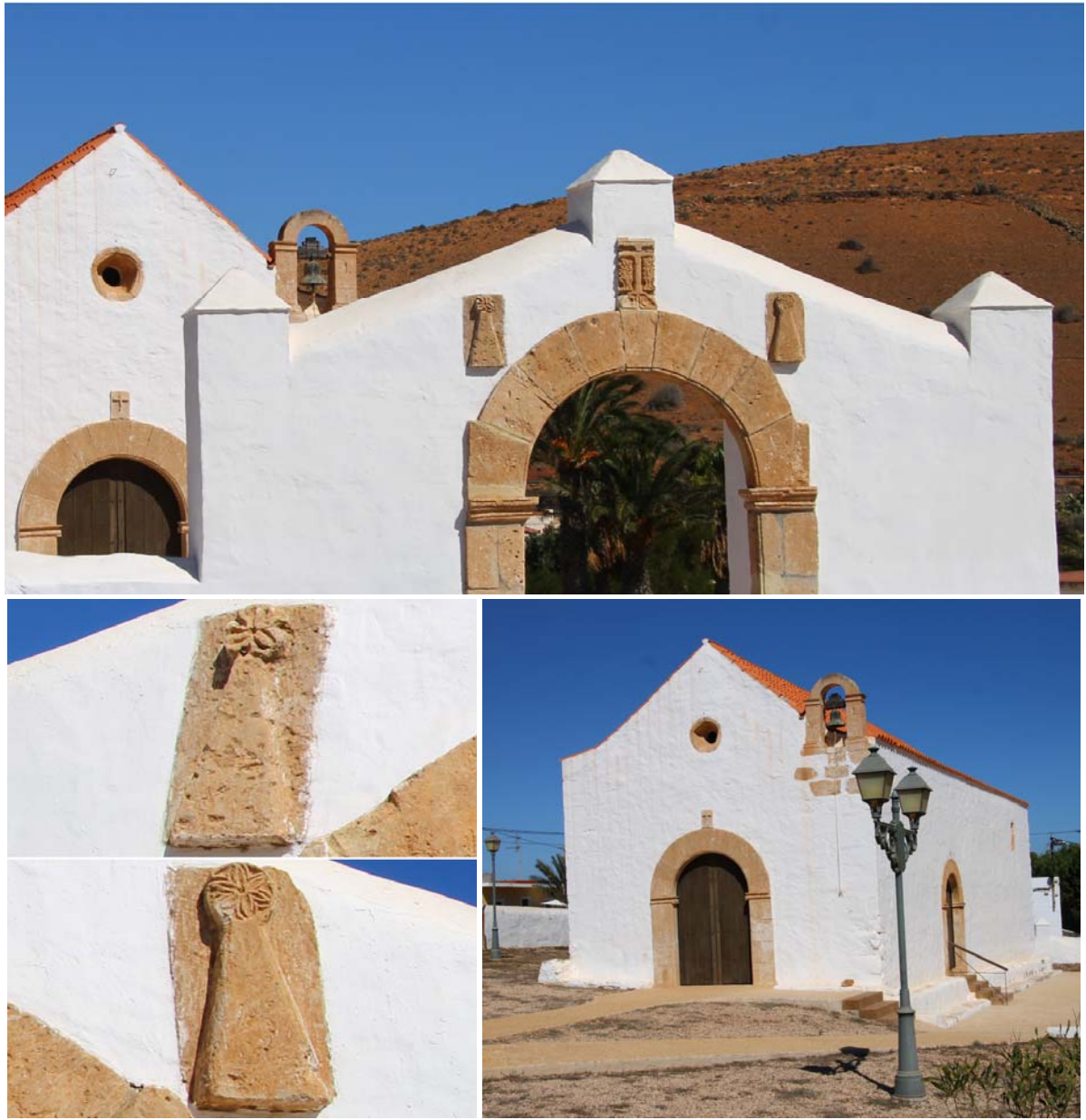

**FIGURE 9.** The simple but beautiful church of Nuestra Señora de Guadalupe (1642) in Agua de Bueyes. It is surrounded by an ample barbican with a large arched and decorated doorway. On the barbican's façade, two small plaques show 8-pointed and 6-pointed rosettes (bottom left images), which might be interpreted as astral symbols, commonly related to Venus and the Sun, respectively. The right side of the building shows a secondary doorway with a few stairs which, given the orientation of the church, opens towards the south as is frequent in the central sector of Fuerteventura.

We agree in that the Sirius' hypothesis is a very speculative one, in need of a more in-depth ethnographic work, as we know that this is an atypical orientation that is substantially different from what it would be expected in the Iberian Peninsula, or even in the rest of the already surveyed islands of the Canary Archipelago. However, given what we have shown here, the idea that certain churches' orientations could respond to a stellar origin is very suggestive.

Summarising this first analysis, we have presented in some detail four possible explanations underlying the peak of approximately –14º in declination of our plots, the second most significant peak of the distribution. As mentioned above, none of the arguments presented proved to be sufficiently conclusive. This is not strange considering the isolation of the island of Fuerteventura, where alien traditions are difficult to explore, and the available sample is limited.

Regarding the main three-fold peak of the declination curvigram (Fig. 8) centred at c. $\delta = +5°$, slightly to the north of due east, there could be one possible explanation. A not negligible group of churches were not oriented to the ecclesiastical equinox on 21 March, as canonical texts apparently suggest, but to the sunrise on Easter Sunday at the approximate year of the construction of the building. This is indeed one of the most important feast days of Christianity. To test this hypothesis, we consider all churches with declination values some five degrees on both sides of the $\delta = +5°$ peak and check the possible Easter dates matching their declination (see Table 2).

**Table 2:** List of chapels and churches of Fuerteventura with declination $+5\pm5°$. The table reproduces data of Table 1, including a new last column. This column indicates the dates of Easter Sundays on years that are close to the most likely years of construction (taking this as the first mention of the building or the date of its last major reform) when the declination of the Sun is approximately the corresponding declination for each church.[5]

| Location | Name (date) | δ (º) (Altar) | Closest Easter date |
|---|---|---|---|
| (12) La Caldereta | N. Sra. de los Dolores (1796) | 6.9±0.7 | 5th Apr 1795 |
| (13) Casillas del Ángel | Santa Ana (1781) | 8.4±0.7 | 11th Apr 1784 |
| (14) La Lajita | N. Sra. de la Inmaculada Concepción (c. 1995) | 7.4±0.7 | 7th Apr 1996 |
| (15) Tetir | Sto. Domingo de Guzmán (c. 1750) | 4.0±0.7 | 29th Mar 1750 / 30th Mar 1755 |
| (16) Pájara | N. Sra. de Regla (1653) | 5.6±0.7 | 4th Apr 1649 |
| (17) Puerto Lajas | Virgen del Pino (c. 1965) | 3.4±0.7 | 29th Mar 1964 |
| (18) Betancuria | Iglesia Convento de San Buenaventura (1653) | 9.4±0.7 | 13th Apr 1653 |
| (19) La Oliva | Ermita interna, Casa de los coroneles (1742) | 3.1±0.7 | 25th Mar 1742 |
| (20) Valle Santa Inés | Santa Inés (c. 1580) | 1.4±0.7 | 'equinoctial' |
| (21) Tesejerague | San José (c. 1725) | 5.2±0.7 | 1st Apr 1725 |
| (22) Las Pocetas | San Francisco Javier (1771) | –0.4±0.7 | 'equinoctial' |

From Table 2 we can verify that given a church value of declination, it is not difficult to find a Sun's declination close to it but occurring during Easter Sunday in a small range of years around the date of construction. With certain caveats, this may suggest a possible explanation for the main peak of the curvigram: it is just the observable evidence that a group of churches were not oriented to the sunrise on 21 March, aka the ecclesiastical equinox, but during the feast of Easter, which results in an accumulation of orientations a bit north of due east. This situation is of course not exclusive of the Canaries. This has been previously suggested for other groups of churches (Romano 1997; Ali and Cunich 2001; Perez-Valcárcel and Perez Palmero 2019), but only recently proven for the Castilian sector of the Way of Saint James (Urrutia-Aparicio *et al.* 2021a). It is important to note that this tradition could have been imported to Fuerteventura in the 15th century after the conquest and colonisation of the island by the Crown of Castile.

Before ending the discussion, it is worth mentioning that the churches of Fuerteventura do not apparently reflect pre-Hispanic traditions in their orientations. The summer solstice was the most important festival for the *Maxos*, the native people of the island, with a concept similar to the equinox in the second term (Perera Betancort *et al.* 1996; Belmonte, 2015). The former is not present in our sample while the latter can easily be confused and would be indistinguishable with the church equinox at 21 March. Therefore, results are inconclusive on this particular aspect.

---

[5] Easter dates from 1600 onwards can be found at: https://www.census.gov/data/software/x13as/genhol/easter-dates.html, accessed October 2021.

## Conclusion

The historic chapels and churches of the Canary Island of Fuerteventura for the most part have orientations that fit within the solar range. Of the 48 measured churches, there are 34 (roughly the 70% of the sample) whose orientations fall within the solar arc, either to the east or to the west. And this proportion increases to 91% if we just consider buildings prior to the 20th century. The rest, which are oriented mainly in the northern and southern quadrants and close to the meridian of the different sites, are, most of them, of recent or private construction.

As we mentioned in the foregoing discussion, this orientation pattern is an expected one according to tradition and to the writings of early Christian authorities. However, more "practical" explanations cannot a priori be excluded, as our analysis also showed that a large proportion of these "canonical" churches possesses a secondary lateral door opening towards the south, in the direction where the annoying north winds cannot enter the church. This is supported by local tradition, as researched by one of the authors, who is native from the island. This differentiates the Christian temples of Fuerteventura from those of the neighbouring island of Lanzarote, where instead of making the faithful access the building through a lateral door, churches were often oriented with their apses towards north-northeast or were provided with prominent barbicans to protect them from the trade winds.

Our results also showed that there is a group of historic churches, mainly located in the central part of Fuerteventura, whose orientations are responsible for an unusual pattern, and which we believe could be related to either one of four following explanations: an orientation towards Easter sunset, the traditional Canarian celebration of "Los Finaos" at the beginning of November, a topographic orientation along the central valleys of the island, and lastly, "la Gañanera". We considered all four hypotheses in some detail but found none to be conclusive.

We also found evidence suggesting that some of the churches explored on the island were oriented towards sunrises – in the different years of their construction, of course – during the wandering feast of Easter, one of the most important festivities of Christianity. However, they do not represent the majority. As noted above, this fact was highlighted by a distinctive accumulation of orientations slightly to the north of due east. Bearing in mind that this tradition of orientations was already found in medieval churches in the Castilian region of the Iberian Peninsula and given that Fuerteventura was colonised by the Crown of Castile, this result may come as no surprise.

In the future, these studies ought to be continued with the measurement of churches' orientations on the few Canary Islands that have yet to be explored, for example the island of Gran Canaria, whose collection of historic churches totals several dozens. Until this archaeoastronomical project is completed in full we will only have a partial understanding of the extent to which canonical orientations were followed in colonial territories located far from the metropolis, which were gradually acquiring a certain degree of autonomy in social and religious matters.

## Acknowledgements

We sincerely acknowledge the incisive commentaries by two anonymous referees who have greatly improved the quality of the paper. The authors wish to thank Joaquín Pallarés and the members of the Agrupación Astronómica de Fuerteventura (AAF) for enlightening discussions on these topics and for their support and kindness during the fieldwork. This work has been partially funded by the following Argentinian institutions: the National Scientific and Technical Research Council (CONICET), under grants reference PIP 11220170100220CO and PIP 11220210100111CO, the University of Buenos Aires (UBACYT 20020190100160BA) and the National University of Luján; and by the State Research Agency (AEI), the Spanish Ministry of Science, Innovation and Universities (MICIU) and the European Regional Development Fund (ERDF), under grants with references AYA2015-66787-P 'Orientatio ad Sidera IV', PID2020-115940GB-C21 'Orientation ad

Sidera V', and under the internal IAC project P310793 'Arqueoastronomía'. M.F.M. is a doctoral fellow of CONICET.